\documentclass[12pt,preprint]{aastex}
\usepackage{natbib}
\usepackage[onecolumn,numberedappendix]{emulateapj5}
\usepackage{epsfig}
\usepackage{rotating}
\tightenlines

\newcommand \lsim{\mathrel{\rlap{\lower4pt\hbox{\hskip1pt$\sim$}}
    \raise1pt\hbox{$<$}}}
\newcommand \gsim{\mathrel{\rlap{\lower4pt\hbox{\hskip1pt$\sim$}}
    \raise1pt\hbox{$>$}}}

\newcommand{\beq}{\begin{equation}}
\newcommand{\eeq}{\end{equation}}
\newcommand{\beqa}{\begin{eqnarray}}
\newcommand{\eeqa}{\end{eqnarray}}

\newlength{\figwidth}
\countdef\decade=200
\advance\decade by \year
\countdef\hours=201
\advance\hours by \time
\divide\hours by 60
\countdef\mins=202
\advance\mins by \hours
\multiply\mins by 60
\multiply\hours by 100
\countdef\miltime=203
\advance\miltime by \hours
\advance\miltime by \time
\advance\miltime by -\mins

\begin{document}

\title{A Test of Star Formation Laws in Disk Galaxies}


\author{Jonathan C. Tan$^1$}
\affil{$^1$Dept. of Astronomy, University of Florida, Gainesville, Florida 32611, USA\\jt@astro.ufl.edu}

\begin{abstract}
We use observations of the radial profiles of the mass surface density
of total, $\Sigma_g$, and molecular, $\Sigma_{\rm H2}$, gas, rotation
velocity and star formation rate surface density, $\Sigma_{\rm sfr}$,
of the molecular dominated regions of 12 disk galaxies from Leroy et
al. to test several star formation laws: a ``Kennicutt-Schmidt power law'',
$\Sigma_{\rm sfr}=A_g \Sigma_{g,2}^{1.5}$; a ``Constant molecular law'',
$\Sigma_{\rm sfr} = A_{\rm H2} \Sigma_{\rm H2,2}$; the
``Turbulence-regulated laws'' of Krumholz \& McKee (KM) and Krumholz,
McKee \& Tumlinson (KMT), a ``Gas-$\Omega$ law'', $\Sigma_{\rm
  sfr} = B_\Omega \Sigma_g \Omega$; and a shear-driven ``GMC
collisions law'', $\Sigma_{\rm sfr} = B_{\rm CC} \Sigma_g \Omega (1 -
0.7 \beta)$, where $\beta \equiv d\:{\rm ln}\:v_{\rm circ} / d\:{\rm
  ln}\: r$.  We find the constant molecular law, KMT turbulence law
and GMC collision law are the most accurate, with an rms error of a
factor of 1.5 if the normalization constants are allowed to vary
between galaxies. Of these three laws, the GMC collision law does not
require a change in physics to account for the full range of star
formation activity seen from normal galaxies to circumnuclear
starbursts. A single global GMC collision law with $B_{\rm
  CC}=8.0\times 10^{-3}$, i.e. a gas consumption time of 20 orbital
times for $\beta=0$, yields an rms error of a factor of 1.8.
\end{abstract}


\section{Introduction}\label{S:intro}

Understanding the rate at which stars form from gas is of fundamental
importance for a theory of galaxy evolution. At the moment it
is uncertain what physical process or processes drive star
formation rates (SFRs). Locally, we know star formation occurs mostly in
highly clustered, $\sim$parsec-scale regions within giant molecular
clouds (GMCs) (Lada \& Lada 2003; Gutermuth et al. 2009). This
clustered mode appears to also be important in a wide range of
galactic environments, including dwarf irregular galaxies (Dowell,
Buckalew, \& Tan 2008), normal disk galaxies (Larsen 2009), and
starburst galaxies (Fall et al. 2005; McCrady \& Graham 2007). The
total efficiency, $\epsilon$, of conversion of gas into stars in these
clusters is relatively high, with $\epsilon \sim 0.1 - 0.5$. However,
on the scale of GMCs star formation occurs at a relatively slow,
inefficient rate, such that only a few percent of the GMC mass is
converted to stars per free-fall time (Zuckerman \& Evans 1974;
Krumholz \& Tan 2007). Although GMCs appear to be gravitationally
bound and approximately virialized (Solomon et al. 1987; Bolatto et al. 2008), at
any given time, most of the mass and volume of GMCs is not forming
stars, perhaps because it is magnetically subcritical (e.g. Heyer et
al. 2008).

Starting with the pioneering work of Schmidt (1959, 1963), empirical
correlations have been found between the disk plane surface density of SFR,
$\Sigma_{\rm sfr}$, and the surface density of gas --- either the total,
$\Sigma_g$, or just that in the molecular phase, $\Sigma_{\rm H2}$.  Based
on about 100 disk averages of nearby galaxies and circumnuclear
starbursts, Kennicutt (1998, hereafter K1998) found
\begin{equation}
\label{sfr1}
\Sigma_{\rm sfr} = A_{g} \Sigma_{g,2}^{\alpha_{g}},
\end{equation}
with $A_g= 0.158 \pm 0.044 \:M_\odot\:{\rm yr^{-1}\:kpc^{-2}}$,
$\Sigma_{g,2} = \Sigma_g / 100 M_\odot {\rm pc^{-2}}$, and $\alpha_{g}
= 1.4\pm0.15$. Most of the dynamic range determining this relation
covers the molecular dominated conditions of the disks in the centers
of normal galaxies and in starbursts. Kennicutt et al. (2007) found a
similar relation applied on $\sim$kpc scales in M51a. Theoretical and
numerical models that relate the SFR to the growth rate of large scale
gravitational instabilities in a disk predict $\alpha_{g}\simeq 1.5$
(e.g. Larson 1988; Elmegreen 1994, 2002; Wang \& Silk 1994; Li, Mac
Low, \& Klessen 2006), as long as the gas scale height does not vary much
from galaxy to galaxy. However, the growth rate of large scale
instabilities that lead to the formation of GMCs cannot be the rate
limiting step for star formation in disks that already have most of
their gas mass in the molecular phase in the form of gravitationally
bound GMCs. Rather, one should consider the processes that create the
actively star-forming, presumably magnetically supercritical,
parsec-scale {\it clumps} of gas within GMCs, which then become star
clusters.

Based on a study of 12 nearby disk galaxies at 800~pc resolution,
Leroy et al. (2008) (see also Bigiel et al. 2008) concluded that
\begin{equation}
\label{sfrH2}
\Sigma_{\rm sfr} = A_{\rm H2} \Sigma_{\rm H2,2},
\end{equation}
with $A_{\rm H2}=(5.25 \pm 2.5) \times 10^{-2}\:M_\odot {\rm
  yr^{-1}\:kpc^{-2}}$ and $\Sigma_{\rm H2,2} = \Sigma_{\rm H2} / 100
M_\odot {\rm pc^{-2}}$. The values of $\Sigma_{\rm H2}$ covered a
range from $\sim 4 - 100 M_\odot\:{\rm pc}^{-2}$. Such a law has also
been proposed as a component of the ``pressure-regulated'' model of
Blitz \& Rosolowsky (2006) in which pressure sets the molecular gas
fraction and then molecular gas forms stars with constant
efficiency. A law similar to eq.~(\ref{sfrH2}) has been implemented by
Shetty \& Ostriker (2008) and, for volume densities, by Kravtsov
(2003). Leroy et al. suggest these results indicate that GMCs in these
galaxies have approximately uniform properties, e.g. density, and thus
are forming stars at a constant rate per free-fall time, as is
expected if they are supersonically turbulent (Krumholz \& McKee 2005,
hereafter KM2005). Such a law was implemented in the simulations of
Dobbs \& Pringle (2009), which are able to reproduce the results of
Bigiel et al. (2008) and Leroy et al. (2008). However, to explain the
K1998 data for higher $\Sigma_g$ systems would require a change in the
cloud properties to allow them to form stars at a faster rate.

KM2005 extended their model of turbulence-regulated
star formation to predict galactic star formation rates by assuming
GMCs are virialized and that their surfaces are in pressure
equilibrium with the large scale interstellar medium (ISM) pressure of
a Toomre (1964) $Q\simeq 1.5$ disk, predicting
\begin{equation}
\label{sfrKM}
\Sigma_{\rm sfr} = A_{\rm KM} f_{\rm GMC} \phi_{\bar{P},6}^{0.34} Q_{1.5}^{-1.32} \Omega_0^{1.32} \Sigma_{g,2}^{0.68},
\end{equation}
with $A_{\rm KM}=9.5 M_\odot \: {\rm yr^{-1}\:kpc^{-2}}$, $f_{\rm
  GMC}$ the mass fraction of gas in GMCs, $\phi_{\bar{P},6}$ the ratio
of the mean pressure in a GMC to the surface pressure here normalized
to a fiducial value of 6 but estimated to vary as
$\phi_{\bar{P}}=10-8f_{\rm GMC}$, $Q_{1.5}=Q/1.5$, and $\Omega_0$
being $\Omega$, the orbital angular frequency, in units of $\rm
Myr^{-1}$. We will assume $f_{\rm GMC}= \Sigma_{\rm H2}/\Sigma_g$
based on resolved studies of GMC populations and molecular gas content
in the Milky Way and nearby galaxies (Solomon et al. 1987; Blitz et al. 2007).

Krumholz, McKee \& Tumlinson (2009a, hereafter KMT2009) presented a two
component star formation law
\begin{eqnarray}
\label{sfrKMT}
\Sigma_{\rm sfr}  =  A_{\rm KMT} f_{\rm GMC} \Sigma_{g,2} \times \left\{ 
\begin{array}{lc}
\left(\Sigma_g/85 M_\odot {\rm pc^{-2}}\right)^{-0.33}, & \Sigma_g< 85\:M_\odot {\rm pc^{-2}}\\
\left(\Sigma_g/85 M_\odot {\rm pc^{-2}}\right)^{0.33}, & \Sigma_g> 85\:M_\odot {\rm pc^{-2}}
\end{array}
\right\}
\end{eqnarray}
with $A_{\rm KMT}=3.85 \times 10^{-2}\:M_\odot {\rm
  yr^{-1}\:kpc^{-2}}$. GMCs are assumed to be in pressure equilibrium
with the ISM only in the high $\Sigma_g$ regime. In the low regime, GMCs
are assumed to have constant internal pressures set by \ion{H}{2}
region feedback (Matzner 2002).

K1998 showed that, in addition to being fit by eq. (\ref{sfr1}), his galaxy and circumnuclear starburst data could be just as well described by
\begin{equation}
\label{sfromega}
\Sigma_{\rm sfr} =  B_{\rm \Omega} \Sigma_g \Omega
\end{equation}
where $B_{\Omega}=0.017$ and $\Omega$ is evaluated at the outer radius
that is used to perform the disk averages. Equation (\ref{sfromega})
implies that a fixed fraction, about 10\%, of the gas is turned into
stars every outer orbital timescale of the star-forming disk and
motivates theoretical models that relate star formation activity to
the dynamics of galactic disks. Such models are appealing as their
predicted star formation activity per unit gas mass, i.e. the gas
consumption time, is self-similar, depending only on the local orbital
time. Examples of these models include those in which star formation
is triggered by passage of gas through spiral density waves (e.g. Wyse
\& Silk 1989).  However, there is no evidence that galactic SFRs
depend on density wave amplitude (e.g. Kennicutt 1989). Rather, where
present, density waves appear to simply help organize gas and star
formation within a galaxy.

Noting that in the main star-forming parts of galactic disks a large
fraction of total gas is associated with gravitationally bound GMCs
and that most stars form in clustered regions in these clouds, Tan
(2000, hereafter T2000) proposed a model of star formation triggered by
GMC collisions in a shearing disk, which reproduces
eq. (\ref{sfromega}) in the limit of a flat rotation curve since the
collision time is found to be a short and approximately constant
fraction, $\sim 20\%$, of the orbital time, $t_{\rm orbit}$. The
collision times of GMCs in the numerical simulations of Tasker \& Tan
(2009) confirm these results. The T2000 model assumes a Toomre $Q$
parameter of order unity in the star-forming part of the disk, a
significant fraction (e.g. $\sim 1/2$) of total gas in gravitationally
bound clouds, and a velocity dispersion of these clouds set by
gravitational scattering (Gammie et al. 1991). Then, the predicted SFR
is
\begin{equation}
\label{sfrcoll}
\Sigma_{\rm sfr} = B_{\rm CC} Q^{-1} \Sigma_g \Omega (1 - 0.7 \beta), \:\:\: (\beta \ll 1)
\end{equation}
where $\beta\equiv d\:{\rm ln}\:v_{\rm circ} / d\:{\rm ln}\: r$ and 
$v_{\rm circ}$ is the circular velocity at a particular galactocentric
radius $r$. Note $\beta=0$ for a flat rotation curve. There is a
prediction of reduced SFRs compared to eq. (\ref{sfromega}) in regions
with reduced shear, i.e. typically the inner parts of disk galaxies.

Leroy et al. (2008) (see also Wong \& Blitz 2002; Bigiel et al. 2008)
examined the applicability of some of the above star formation laws
for the galaxies in their sample. In this Letter we revisit this
issue, concentrating on the radial profiles of the molecular dominated
regions of the 12 disk galaxies studied by Leroy et al.


\section{Methodology}\label{S:method}

\begin{deluxetable}{cccc|cc|cc|cc|cc|cc|cc|}
\tabletypesize{\footnotesize}
\tablecolumns{16}
\tablewidth{0pt}
\tablecaption{Star Formation Law Parameters for Sample Galaxies}
\tablehead{\colhead{Galaxy} &
           \colhead{d} &
           \colhead{$r_{\rm out}$} &
           \colhead{$N_{\rm ann}$} &
           \colhead{$A_{g}$\tablenotemark{a}} &
           \colhead{$\chi_{g}$} &
           \colhead{$A_{\rm H2}$\tablenotemark{a}} &
           \colhead{$\chi_{\rm H2}$} &
           \colhead{$A_{\rm KM}$\tablenotemark{a}} &
           \colhead{$\chi_{\rm KM}$} &
           \colhead{$A_{\rm KMT}$\tablenotemark{a}} &
           \colhead{$\chi_{\rm KMT}$} &
           \colhead{$B_{\rm \Omega}$} &
           \colhead{$\chi_{\Omega}$} &
           \colhead{$B_{\rm CC}$} &
           \colhead{$\chi_{\rm CC}$} \\
	   \colhead{NGC:} &
	   \colhead{\tiny (Mpc)} &
	   \colhead{\tiny (kpc)} &
	   \colhead{} &
	   \colhead{\tiny($10^{-2}$)} &
	   \colhead{\tiny($10^{-2}$)} &
	   \colhead{\tiny($10^{-2}$)} &
	   \colhead{\tiny($10^{-2}$)} &
	   \colhead{} &
	   \colhead{\tiny($10^{-2}$)} &
	   \colhead{\tiny($10^{-2}$)} &
	   \colhead{\tiny($10^{-2}$)} &
	   \colhead{\tiny($10^{-3}$)} &
	   \colhead{\tiny($10^{-2}$)} &
	   \colhead{\tiny($10^{-3}$)} &
	   \colhead{\tiny($10^{-2}$)} 
}
\startdata
628 & 7.3 & 3.7 & 11 & 11.0 & 6.71 & 6.41 & 11.8 & 0.750 & 32.1 & 3.75 & 10.0 & 4.05 & 22.9 & 5.37 & 11.8\\
2841 & 14.1 & 7.9 & 7 & 14.0 & 8.72 & 5.50 & 7.53 & 1.07 & 16.2 & 2.22 & 5.83 & 6.00 & 8.39 & 6.01 & 8.28 \\
3184 & 11.1 & 5.1 & 10 & 7.43 & 4.40 & 4.63 & 9.87 & 1.60 & 15.5 & 2.80 & 12.1 & 6.43 & 9.68 & 11.7 & 11.7\\
3198 & 13.8 & 1.7 & 3 & 27.0 & 10.1 & 14.9 & 8.58 & 4.82 & 17.6 & 7.69 & 18.6 & 20.2 & 19.6 & 48.3 & 26.5\\
3351 & 10.1 & 4.7 & 10 & 23.4 & 27.1 & 10.6 & 14.4 & 1.57 & 7.14 & 5.12 & 24.8 & 8.85 & 8.67 & 10.9 & 13.3\\
3521 & 10.7 & 6.5 & 13 & 5.19 & 4.56 & 4.22 & 1.97 & 1.25 & 20.6 & 3.14 & 5.19 & 4.64 & 13.0 & 6.26 & 2.11\\
3627 & 9.3 & 7.4 & 17 & 8.32 & 34.2 & 5.77 & 25.3 & 2.08 & 35.4 & 3.77 & 24.9 & 7.90 & 36.8 & 9.90 & 25.7\\
4736 & 4.7 & 1.7 & 8 & 14.8 & 21.8 & 13.8 & 20.5 & 1.02 & 40.9 & 10.8 & 10.2 & 4.82 & 33.8 & 5.34 & 27.2\\
5055 & 10.1 & 8.6 & 18 & 4.88 & 16.3 & 3.92 & 10.5 & 1.51 & 34.9 & 2.76 & 8.88 & 5.22 & 29.9 & 5.82 & 21.1\\
5194 & 8.0 & 6.0 & 16 & 5.79 & 28.7 & 5.84 & 25.8 & 1.71 & 36.7 & 4.50 & 20.9 & 5.64 & 33.1 & 6.76 & 26.1\\
6946 & 5.9 & 5.9 & 21 & 6.88 & 29.1 & 6.45 & 19.5 & 2.59 & 29.0 & 4.73 & 17.5 & 8.31 & 29.4 & 11.5& 18.2\\
7331 & 14.7 & 6.8 & 10 & 7.47 & 5.61 & 4.93 & 2.74 & 1.26 & 26.7 & 3.31 & 5.09 & 5.39 & 17.2 & 6.98 & 5.14\\
\hline
$N_{\rm fit}$=12 & & & 144 & & 22.1 & & 16.8 & & 29.7 & & 16.1 & & 26.3 & & 18.8\\
$N_{\rm fit}$=1 & & & 144 & 8.17 & 29.4 & 5.95 & 22.2 & 1.57 & 32.8 & 3.92 & 21.8 & 6.24 & 28.4 & 8.05 & 24.9\\
\enddata

\tablenotetext{a}{Units: $M_\odot {\rm yr^{-1}kpc^{-2}}$}

\label{tb:gal}
\end{deluxetable}

We consider the data on $\Sigma_{\rm sfr}$, $\Sigma_g$, $\Sigma_{\rm
  H2}$, $\Omega$ and $\beta$ for the 12 large disk galaxies (see Table
\ref{tb:gal}) analyzed by Leroy et al. (2008), and we refer the reader
to this paper for the details of how these quantities were
estimated. Note that $\Omega$ and $\beta$ depend on the estimated
rotation curves of the galaxies. The Leroy et al. (2008) analysis uses
analytic fits to the observed rotation curves, since the derivatives
of the actual observed curves can be very noisy.

We only consider regions where the molecular gas dominates over
atomic, i.e. $\Sigma_{\rm H2}\geq\Sigma_{\rm HI}$, since it is here that
we expect a significant fraction of the total gas to be associated
with gravitationally bound clouds --- an assumption of the T2000 and
KM2005 theories --- and since we also wish to avoid regions affected
by star formation thresholds (Martin \& Kennicutt 2001). This
requirement defines an outer radius, $r_{\rm out}$, for each
galaxy. Note that NGC 2841 has no detected gas in its central region
out to about 3.5~kpc, so we only consider annuli from this radius out
to $r_{\rm out}$ for this galaxy. The requirement that $\Sigma_{\rm
  H2}>\Sigma_{\rm HI}$ also leads us to exclude analysis of the 11
\ion{H}{1} dominated, low-mass galaxies in the Leroy et al. (2008)
sample, which have only upper limits on $\Sigma_{\rm H2}$.

We use these data to compare the predicted $\Sigma_{\rm sfr,theory}$
from: a ``Kennicutt-Schmidt power law'' with $\alpha_g=1.5$ (eq. \ref{sfr1});
a ``Constant molecular law'' (eq. \ref{sfrH2}); a ``KM2005
turbulence-regulated law'' (eq. \ref{sfrKM}); a ``KMT2009
turbulence-regulated law'' (eq. \ref{sfrKMT}) a ``Gas-$\Omega$ law''
(eq. \ref{sfromega}); and a ``GMC collision law'' (eq. \ref{sfrcoll}),
with the observed values, $\Sigma_{\rm sfr,obs}$, averaged in annuli
of typical width $\sim 500$~pc. For each galaxy and each star
formation law we derive the best fit values of $A_g$, $A_{\rm H2}$,
$A_{\rm turb}$, $B_{\rm \Omega}$ and $B_{\rm CC}$, respectively,
weighting all $N_{\rm ann}$ annuli equally, and we measure the rms
factor by which the theoretical model errs in estimating the SFR as
$10^\chi$, where $\chi^2\equiv (N_{\rm ann} - N_{\rm fit})^{-1} \sum
({\rm log} R_{\rm sfr})^2$, $R_{\rm sfr}= \Sigma_{\rm
  sfr,theory}/\Sigma_{\rm sfr,obs}$ and $N_{\rm fit}=1$. We also
derive values of $\chi$ for the entire galaxy sample for the case
where each galaxy is allowed one free parameter ($N_{\rm fit}=12$) and
for the case where there is a single star formation law with one free
parameter. As discussed by KM2005, these are not traditional $\chi^2$
goodness-of-fit statistics, but the factor $10^\chi$ does give a
measure of how well the theoretical model reproduces the observed
system. Note that the models vary in the number of observables they
depend on and so measurement errors will introduce varying degrees of
dispersion.

\section{Results \& Discussion}\label{S:results}

\begin{figure}[h]
\begin{center}
\epsfig{
        file=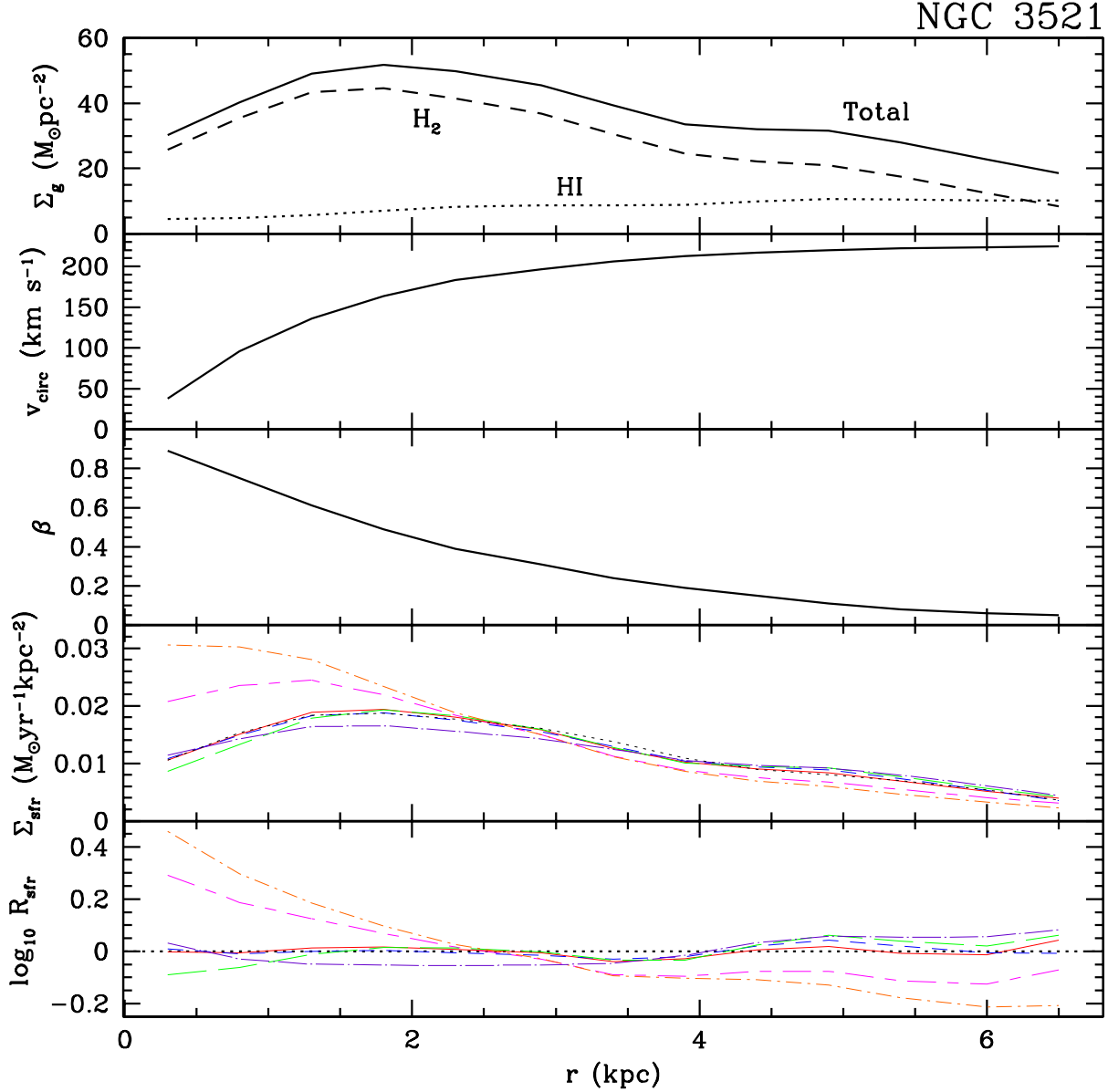,
        angle=0,
        width=6.7in
}
\end{center}
\caption{\label{fig:1} Radial distribution of properties of NGC 3521
  from Leroy et al. (2008) over the region where molecular gas mass
  dominates over atomic. First (top) panel: Mass surface densities of
  total gas, $\Sigma_g$ (solid), molecular phase including He,
  $\Sigma_{\rm H2}$ (dashed), and atomic phase including He,
  $\Sigma_{\rm HI}$ (dotted). Second panel: Circular velocity, $v_{\rm
    circ}$. Third panel: logarithmic derivative of rotation curve,
  $\beta \equiv d {\rm ln}\: v_{\rm circ} / d {\rm ln}\: r$. Fourth
  panel: Star formation rate surface density, $\Sigma_{\rm sfr}$:
  observed (dotted), predicted via $\Sigma_{\rm sfr}=A_g
  \Sigma_g^{1.5}$ (green long-dashed), via $\Sigma_{\rm sfr} = A_{\rm
    H2} \Sigma_{\rm H2,2}$ (blue dashed), via $\Sigma_{\rm sfr} =
  A_{\rm KM} f_{\rm GMC} \phi_{\bar{P},6}^{0.34} Q_{1.5}^{-1.32}
  \Omega_0^{1.32} \Sigma_{g,2}^{0.68}$ (orange dot-dashed), via the
  KMT2009 turbulence law (eq. \ref{sfrKMT}) (purple dot-long-dashed),
  via $\Sigma_{\rm sfr}=B_\Omega \Sigma_g \Omega$ (magenta
  dashed-long-dashed), and via $\Sigma_{\rm sfr}=B_{\rm CC} \Sigma_g
  \Omega (1 - 0.7 \beta)$ (red solid). Fifth (bottom) panel: Ratio,
  $R_{\rm sfr}$, of predicted to observed star formation rate surface
  densities, with line types as in the fourth panel. In this galaxy,
  the constant molecular (blue dashed) and GMC collision (red solid)
  star formation laws are the most accurate at predicting the radial
  profile of $\Sigma_{\rm sfr}$.  }
\end{figure}

Figure~\ref{fig:1} shows an example of our method applied to galaxy
NGC~3521, displaying the observed profiles of molecular, atomic and
total gas content, $v_{\rm circ}$, $\beta$, observed and predicted
$\Sigma_{\rm sfr}$ and $R_{\rm sfr}$. In this galaxy the star
formation laws based on conversion of molecular gas at fixed rate and
based on GMC collisions provide the best fit to the observations. 

\begin{figure}[h]
\begin{center}
\epsfig{
        file=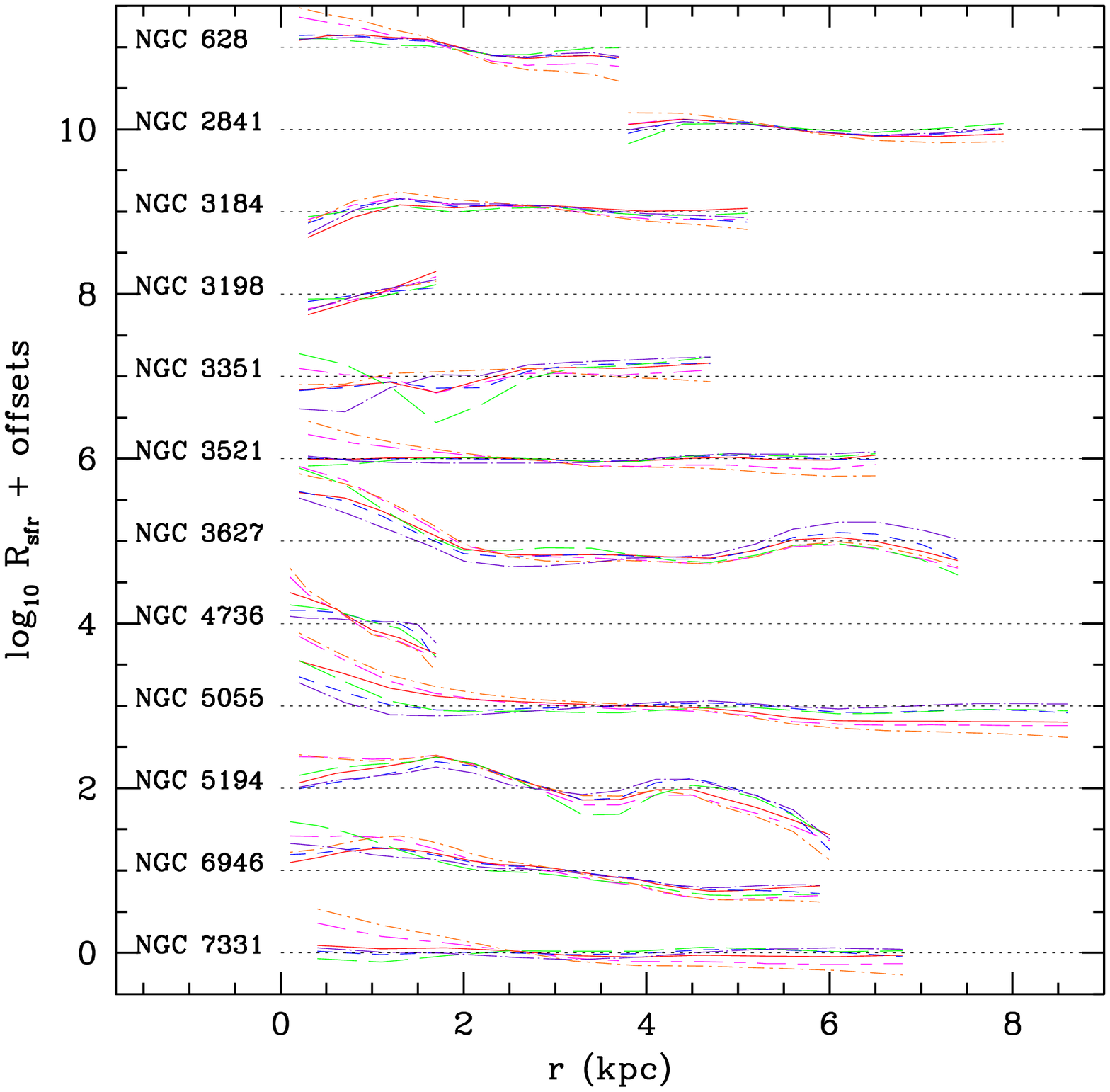,
        angle=0,
        width=6.7in
}
\end{center}
\caption{ \label{fig:2} 
Ratio, $R_{\rm sfr}$, of predicted to observed star formation rate surface densities for the entire sample of disk galaxies, offset from each other for clarity: the dotted lines indicate $R_{\rm sfr}=1$ for each galaxy. The line styles are as in Figure~\ref{fig:1}.
}
\end{figure}

Figure~\ref{fig:2} shows $R_{\rm sfr}$ for the entire sample of disk
galaxies and Table~\ref{tb:gal} lists the best fit parameters of these
models and their dispersions in log~$R_{\rm sfr}$. In order of
decreasing accuracy the models are KMT2009 turbulence-regulated,
constant molecular, GMC collisions, Kennicutt-Schmidt power law,
gas-$\Omega$ law and KM2005 turbulence-regulated, although we caution
that given the systematic uncertainties and the different number of
observables associated with each law some of these distinctions are
probably not significant. Nevertheless, with the freedom to adjust one
parameter for each galaxy, which can account for certain systematics
such as distance uncertainties, global metallicity variations,
foreground extinctions, etc., the KMT2009 turbulence-regulated,
constant molecular and GMC collision models can reproduce the observed
star formation rates with rms errors of factors of 1.5 (factors of
about 1.7 for a single parameter star formation law), while the other
models do somewhat worse, e.g. factors of about 2.0 for the KM2005
turbulence-regulated model. We note that the constant molecular law
would require modification to explain the observed super-linear
dependence of $\Sigma_{\rm sfr}$ on $\Sigma_{g}$, i.e. the
Kennicutt-Schmidt relation, that extends to circumnuclear starbursts
(K1998). Also this law and the turbulence-regulated laws have had the
advantage of using the observed fraction of molecular gas as an
input. If the KMT2009 law uses the molecular fractions predicted from
the observed metallicities (Krumholz, McKee, \& Tumlinson 2009b), the
value of $\chi_{\rm KMT}$ rises to 0.17 and 0.25 for $N_{\rm
  fit}=12,1$, respectively (with $A_{\rm KMT}=0.038$).

In addition to the rms dispersion, we note that there are systematic
trends with $r$: many of the models tend to over predict SFRs in the
galactic centers and under predict in outer regions. This indicates
that a Kennicutt-Schmidt power law with $\alpha_g>1.5$ will do worse
than one with $\alpha_g=1.5$. In the context of the GMC collision law,
the galactic centers are where $\beta$ is relatively large, i.e. of
reduced shear, for which the T2000 model (eq. \ref{sfrcoll}) is not
expected to be particularly accurate. Future numerical simulations,
extending the analysis of Tasker \& Tan (2009), can more accurately
measure the dependence of the GMC collision rate as a function of
$\beta$ to help develop a more refined model. Note also that in the
limit of pure solid body rotation the GMC collision law needs to be
modified to include a mode of star formation not set by shear-driven
collisions, i.e. perhaps regulated by magnetic fields or turbulence.

Any test of SFR laws against observational data is necessarily
limited in scope by being able to test only those laws that utilize
the particular properties that have been observed. The physics
underlying the law that does best in this test is not
necessarily that which controls galactic SFRs. Nevertheless,
comparison between the laws that can be tested does provide evidence
for the relative merit of the physical mechanisms they invoke.

We conclude that a model of star formation controlled by shear-driven
GMC collisions, $\Sigma_{\rm sfr} = 8.0\times 10^{-3} Q^{-1} \Sigma_g
\Omega (1 - 0.7 \beta)$, provides an accurate description of the
radial profiles of $\Sigma_{\rm sfr}$ in the molecular dominated
regions of 12 nearby disk galaxies, performing favorably in comparison
with a range of other models and better than the Kennicutt-Schmidt
power law, the gas-$\Omega$ law and the KM2005 turbulence-regulated
model. Additionally, it is a relatively simple and physically
well-motivated model for explaining SFRs in both disk galaxies and the
more extreme circumnuclear starbursts studied by K1998.
Future high angular resolution studies with ALMA to measure galactic
radial profiles of velocity dispersion, and thus $Q$, will improve our
ability to test the above star formation laws. Such observations will
also be able test laws, such as those involving regulation by
turbulence, that make predictions for the explicit dependence of SFR
with velocity dispersion. 

\acknowledgements We thank Adam Leroy and Mark Krumholz for
discussions and Adam Leroy for providing his published data in a
convenient format. The comments of an anonymous referee helped to
improve the paper. JCT acknowledges support from NSF CAREER grant
AST-0645412.

\end{document}